\documentclass[11pt,a4paper]{article}
\usepackage{amsmath}

\usepackage[psamsfonts]{amssymb}
\usepackage{amsmath}
\usepackage{epsfig}

\author{H. Mohseni Sadjadi\footnote{mohseni@phymail.ut.ac.ir;
mohsenisad@ut.ac.ir}  and  N. Vadood
\\ {\small Department of Physics, University of Tehran ,}
\\ {\small P.O.B. 14395-547, Tehran 14399-55961, Iran}}
\title{ Notes on interacting holographic dark energy model in a closed universe}

\begin{document}
\maketitle
\begin{abstract}
We consider interacting holographic dark energy model in Friedmann
Robertson Walker space time with positive spatial curvature and
investigate the behavior of geometric parameter and dark energy
density in accelerated expanding epoch. We also derive some
conditions needed to cross the phantom divide line in this model.
\end{abstract}

\section{Introduction}

To describe the present acceleration of the universe \cite{acc}
different models have been proposed. If we adopt the Einstein
theory of gravity, this acceleration is only possible when
approximately $70\%$ of the universe is filled with a component
with negative pressure dubbed as dark energy. A straightforward
candidate for dark energy is the vacuum energy which suffers from
conceptual problems such as fine-tuning and coincidence problems
\cite{vac}. The amount of the dark energy density assessed in this
model differs of 120 order of magnitude from the observational
value. Some present data seem to favor a dark energy component
with an equation of state (EoS) parameter, $w_d$, evolving from a
value greater than $-1$ in the past to $w_d<-1$ in the present
epoch \cite{cross}. This dynamical behavior cannot be explained by
the cosmological constant which possesses a constant EoS
parameter: $w_d=-1$. Observations also show that the dark energy
and dark matter densities are of the same order at the present
epoch (known as coincidence problem). This would not be true if
there were no interactions between these components. Indeed in
dark energy models, as the universe expands, the ratio of matter
to dark energy density is expected to decrease rapidly
(proportional to the scale factor). To solve these problems, one
can adopt an evolving dark energy with suitable interaction with
(dark) matter \cite{inter}.

One of the models proposed to describe the present accelerated
expansion of the universe, and the dynamical behavior of EoS
parameter, is the holographic dark energy model
\cite{Li1},\cite{holo}. This model is based upon the fact that the
formation of a black hole requires a relation between the
ultraviolet and infrared cutoffs of the system which leads us to
assume that the total dark energy contained in a system must not
exceed the mass of the black hole of the same size \cite{coh}. In
this way the dark energy density may be related to the dynamical
infrared cutoff of the system \cite{Li1}. Note that besides the
late time acceleration, the holographic dark energy model also may
be used to study the inflationary and post-inflationary epochs of
the universe \cite{inf}.

In this paper we consider interacting holographic dark energy
model \cite {abd} in a Friedmann Robertson Walker space time with
positive spatial curvature. We don't restrict ourselves to only
small curvature limit and discuss time evolution of dark energy
and dark matter densities. We investigate the behavior of
geometric parameter in accelerated expanding epoch. We allow the
dark energy to exchange energy with (dark) matter and discuss
conditions needed to cross phantom divide line, by considering the
thermodynamics second law. We show that at the transition time
there may be an upper limit for dark energy density, which depends
upon the interaction parameters as well as geometric parameter
which may be regarded as geometrical correction (due to departure
from flatness) to our previous flat case results in \cite{me}. Our
results may also alleviate the coincidence problem.

Throughout this paper we use $\hbar=c=G=k_{B}=1$ units.

\section{Holographic dark energy}
\subsection{Properties and Evolution }
We consider Friedmann Robertson Walker (FRW) space time described
by the metric
\begin{equation}\label{1}
ds^2=-dt^2+a^2(t)\left({
dr^2\over{1-kr^2}}+r^2(d\theta^2+sin^2\theta d\phi^2)\right),
\end{equation}
in the comoving coordinates. $a(t)$ is the scale factor and $k$
determines the spatial curvature of the space-time. The universe
is assumed to be filled with perfect fluid(s) at large scale. The
Hubble parameter, $H$, is related to energy density, $\rho$, via
the Friedmann equation
\begin{equation}\label{2}
H^2={8\pi\over 3}\rho-{k\over a^2}.
\end{equation}
We have also the evolution equation
\begin{equation}\label{3}
\dot{H}=-4\pi(P+\rho)+{k\over a^2},
\end{equation}
where $P$ is the pressure. When the total density is equal to the
critical density defined by $\rho_c={3H^2\over 8\pi}$, the
universe is spatially flat, i.e. $k=0$. We assume that the
universe is dominated by pressureless dark matter (denoted by the
subscript $m$) and dark energy component (denoted by the subscript
$d$). In this paper we restrict ourselves to positively curved
space with three dimensional spatial spherical geometry and take
$k=1$. The relative densities defined by $\Omega_m={\rho_m\over
\rho_c}$, and $\Omega_d={\rho_d\over \rho_c}$ satisfy
\begin{equation}\label{4}
\Omega_m+\Omega_d-\Omega_k=1,
\end{equation}
where the geometric parameter, $\Omega_k$, is defined as
$\Omega_k={1\over (aH)^2}$. The equation of state parameter of the
dark energy, $w_d$, given by $P_d=w_d \rho_d$, satisfies
\begin{equation}\label{5}
w_d={w(1+\Omega_k)\over \Omega_d},
\end{equation}
where $w$ is the EoS parameter of the universe. The time evolution
of ${\Omega_k}$ is obtained as
\begin{equation}\label{6}
\dot{\Omega_k}=-2H\Omega_k(1+{\dot{H}\over H^2}).
\end{equation}
We consider a model of dark energy and dark matter interacting via
the source term $(\lambda_m\rho_m+\lambda_d\rho_d)H$. So there is
energy exchange between dark matter and dark energy components.
While these components are not conserved,
\begin{eqnarray}\label{7}
\dot{\rho_d}+3H\rho_d(1+w_d)&=&-(\lambda_m\rho_m+\lambda_d\rho_d)H \nonumber \\
\dot{\rho_m}+3H\rho_m&=&(\lambda_m\rho_m+\lambda_d\rho_d)H,
\end{eqnarray}
the total density satisfies the continuity equation
\begin{equation}\label{8}
\dot{\rho}+3H\rho(1+w)=0.
\end{equation}
To study how the ratio of $\Omega_m$ to $\Omega_d$ changes with
time, one can use
\begin{equation}\label{9}
\dot{r}=H(1+r)\big((\lambda_m +3w)r+\lambda_d\big),
\end{equation}
where
\begin{equation}\nonumber
r={\rho_m\over {\rho_d}}={\Omega_m\over {\Omega_d}}.
\end{equation}
In the absence of interaction ($\lambda_m=\lambda_d=0$), $r$ is a
decreasing (increasing) function of time, when $w<0(>0)$. But in
the presence of interaction $r$ may be increasing even in
accelerating phase.

The time derivative of the ratio $\mathcal{P}:={\Omega_k\over
\Omega_d}={3\over 8\pi}{1\over a^2\rho_d}$, has the same sign as
$(1+3w_d+\lambda_d)+\lambda_mr$. To verify this claim one can use
({\ref{7}}). When the components are non-interacting and in
(non-)accelerating phase, $w_d>(<){-{1\over 3}}$, we have
$\dot{\mathcal{P}}>(<)0$. In the presence of interaction this
claim is not generally true and the behavior of $\mathcal{P}$
depends upon the interactions and conditions considered in the
model.

We take the dark energy component as a holographic dark energy
determined through
\begin{equation}\label{10}
\rho_d={3c^2\over 8\pi L^2},
\end{equation}
where $c$ is a numerical constant and $L$ is an infrared cutoff
which may be chosen as follows. Assume a  light signal which is
emitted from ${\bold r}$ at $t$ will arrive at the origin at
$t=\infty$, as the light signal propagates along the geodesic
$ds^2=0$, we have
\begin{equation}\label{11}
\int_t^\infty {dt \over a(t)}=\int_0^{\bold r}{dr\over
\sqrt{1-r^2}}=\sin^{-1}{\bold r}.
\end{equation}
We choose $L$ as the radius of the event horizon measured on the
sphere of the horizon (see the second reference in \cite{Li1}) ,
hence $L=a(t){\bold r}$. Defining $R_h=a(t)\int_t^\infty {dt\over
a(t)}$, we obtain $L=a(t)\sin y$, where $y={R_h\over a(t)}$. In
the flat case, $k=0$, and $L$ reduces to $L=R_h=a(t)\int_t^\infty
{dt\over a(t)}$. One can assign an entropy to the universe
characterized by the cutoff $L$ as
\begin{equation}\label{12}
S=\pi L^2.
\end{equation}
The time derivative of $L$ can be shown to be
\begin{equation}\label{13}
\dot{L}=HL-\cos y.
\end{equation}
Hence the thermodynamics second law, $\dot{S}\geq 0$, is valid
whenever
\begin{equation}\label{14}
0<{\Omega_d^{1\over 2}\over c} \cos y\leq 1,
\end{equation}
or in terms of $\Omega_k$
\begin{equation}\label{15}
\Omega_d\leq c^2(1+\Omega_k).
\end{equation}
Note that $\ddot{L}=-{L\over a^2}+(HL\dot{)}$. For
$\dot{\Omega_d}>0$, we have $(HL\dot{)} < 0$ which leads to
$\ddot{L} < 0$. But if one requires that the entropy attributed to
the cutoff $L$ is increasing, he finds $\dot{L}>0$, then either
$\lim_{t\to \infty}\dot{L}= 0$ or $\ddot{L}$ becomes positive
after a finite time, i.e., $\dot{\Omega_d}>0$ will no more be
valid.

The equation of state parameter of the compact universe is
\begin{equation}\label{16}
w=-1-{2\over 3}{{{\dot{H}\over H^2}-\Omega_k}\over {1+\Omega_k}},
\end{equation}
which leads to $1+{\dot{H}\over H^2}=-{1\over
2}(1+\Omega_k)(1+3w)$. Therefore like the flat case, we have
$\ddot{a}>0$ when $w<-{1\over 3}$. If $\Omega_k\neq 0$, we obtain
$w=-{1\over 3}+{\dot{\Omega_k}\over{3H\Omega_k(1+\Omega_k)}}$.
Thus the sign of $\dot{\Omega_k}$ determines whether the universe
is in accelerated phase ($w<-{1\over 3}$) or not. The super
accelerated universe $\dot{H}>0$ corresponds to $w<-{1\over
3}\left(1+{2\over {1+\Omega_k}}\right)$.

Taking time derivative of both sides of $HL=c\Omega_d^{-{1\over
2}}$ (which may be derived from (\ref{10})) leads to: $\dot{H}
L+H^2L+{c\over 2}\Omega_d^{-{3\over 2}}\dot{\Omega_d}=H\cos y$,
therefore from (\ref{16}) we obtain
\begin{equation}\label{17}
w=-{1\over 3}-{2\cos y\over 3c(1+\Omega_k)}\Omega_d^{1\over
2}+{1\over 3H(1+\Omega_k)}{\dot{\Omega_d}\over \Omega_d}
\end{equation}
For $w<-{1\over 3}$, from the above equation we deduce
$\dot{\Omega_d}\leq {2\over c}H\Omega_d^{3\over 2}\cos y $, which
by considering the thermodynamics second law results in
$\dot{\Omega_d}\leq 2H\Omega_d$, implying $\dot{\Omega_d}\leq
2H(1+\Omega_k)$. It can be shown that
\begin{equation}\label{18}
\dot{r}={\dot{\Omega_k}\over
\Omega_d}-\dot{\Omega_d}{1+\Omega_k\over \Omega_d^2}.
\end{equation}
Comparing this result with (\ref{9}) yields
\begin{equation}\label{19}
w={\dot{\Omega_k}\Omega_d\over
3H\Omega_m(1+\Omega_k)}-{\dot{\Omega_d}\over 3H\Omega_m}-{1\over
3}(\lambda_m+{\lambda_d\Omega_d\over \Omega_m}).
\end{equation}
Using (\ref{17}) and (\ref{19}) and
\begin{equation}\label{24}
\dot{\Omega_k}=H\Omega_k(1+3w)(1+\Omega_k),
\end{equation}
$w$ and $\dot{\Omega_d}$ may be obtained as
\begin{eqnarray}\label{20}
w&=&-{2\cos y\over 3c(1+\Omega_k)}\Omega_d^{3\over
2}+{\lambda_m-\lambda_d-1\over
3(1+\Omega_k)}\Omega_d-{\lambda_m\over 3},\nonumber \\
{\dot{\Omega_d}\over H}&=&-{1+\lambda_d-\lambda_m\over
1+\Omega_k}\Omega_d^2+{2\over c}\left(1-{\Omega_d\over 1+\Omega_k}
\right)\Omega_d^{3\over 2}\cos{y}+(1-\lambda_m)\Omega_d\nonumber
\\
&&+\Omega_d\Omega_k(1+3w).
\end{eqnarray}
Note that study of this model is more complicated with respect to
the flat case where the right hand side of the above equation (
besides $\lambda_m$, $\lambda_d$ and $c$) depends only on
$\Omega_d$ \cite{me}:
\begin{eqnarray}\label{40}
w&=&-{2\Omega_d^{3\over 2}\over 3c}+{\lambda_m-\lambda_d-1\over
3}\Omega_d-{\lambda_m\over 3},\nonumber \\
{\dot{\Omega_d}\over H}&=&\Omega_d({2\over c}\Omega_d^{1\over
2}+3w+1).
\end{eqnarray}
By (\ref{20}) and (\ref{14}), we obtain
$(3w+\lambda_m)(r+1)+3+\lambda_d-\lambda_m\geq 0$ or
\begin{equation}\label{21}
{\dot{r}\over H(1+r)}\geq -3(1+w).
\end{equation}
Hence if $w<-1$ then $\dot{r}>0$ indicating that the ratio of dark
matter to dark energy increases. For $w<-{1\over 3}$, we obtain
$\dot{r}\geq -2H(1+r)$.  The evolution of the ratio of $\Omega_k$
to $\Omega_d$, represented by $\mathcal{P}$ can be given by
\begin{eqnarray}\label{22}
\dot{\mathcal{P}}&=&{3H\over
8\pi}{(1+3w_d+\lambda_d)\rho_d+\lambda_m\rho_m\over
a^2\rho_d^2}\nonumber \\
&=&(1+3w_d+\lambda_d+\lambda_m r)\mathcal{P}\nonumber \\
&=&-{2\over c}H\mathcal{P}\Omega_d^{1\over 2}\cos y.
\end{eqnarray}
Hence if the thermodynamics second law in the form (\ref{14}) is
valid then $\mathcal{P}$ must be a decreasing function of time.

\subsection{ $w=-1$ crossing} In order that the effective EoS
parameter crosses $w=-1$, we must have $w_d<-1-r$, which requires
$\Omega_m<-(1+w_d)\Omega_d$ or $\Omega_k<-1-w_d\Omega_d$. If the
transition is assumed to be from quintessence to phantom phase,
then $\dot{w}$ must be negative at $w=-1$. From (\ref{20}) we have
\begin{eqnarray}\label{23}
&&\dot{w}={\dot{\Omega_d}\over {1+\Omega_k}}\left(-{1\over
c}\Omega_d^{1\over 2}\cos y-{c\over 3
\cos{y}}\Omega_k\Omega_d^{-{1\over
2}}-{1\over 3}(1+\lambda_d-\lambda_m)\right)\nonumber \\
&&+{\dot{\Omega_k}\over (1+\Omega_k)^2}\Big( {2\over
3c}\Omega_d^{3\over 2}\cos y+{1\over
3}(1+\lambda_d-\lambda_m)\Omega_d\nonumber \\
&&+{c\over 3\cos y}(1+\Omega_k)\Omega_d^{1\over 2} \Big).
\end{eqnarray}
Using (\ref{24}), (\ref{23}) becomes
\begin{equation}\label{25}
\dot{w}=-{2H\Omega_d\over
1+\Omega_k}\left[X^2+\left({\lambda_d-\lambda_m+1\over
3}+{\alpha\over 6}(\Omega_k+3)\right)X+{\alpha\over
6}(\lambda_d-\lambda_m+1)+{\Omega_k\over 3}\right],
\end{equation}
where  $X={1\over c}\Omega_d^{1\over 2}\cos y$, and $\alpha=1+3w$.
At $w=-1$, (\ref{25}) reduces to
\begin{equation}\label{26}
\dot{w}=-{2H\Omega_d\over 1+\Omega_k}[(X-1)(X+{1\over
3}(\lambda_d-\lambda_m+1-\Omega_k))].
\end{equation}
Thermodynamics second law implies that $X\leq 1$. For $X=1$, we
obtain  $\dot{w}=0$ at $w=-1$. But
\begin{equation}\label{27}
\dot{X}=H[X^2+{1\over 2}(1+\Omega_k)(1+3w)X+\Omega_k],
\end{equation}
therefore if $X=1$ at $w=-1$, then we also must have $\dot{X}=0$.
In the same way, using (\ref{25}) one can show that ${d^n X\over
dt^n}=0$ which results in ${d^n w\over dt^n}=0$ at $w=-1$. Hence
$X=1$ at $w=-1$ implies that $\dot{X}$, and higher derivatives of
$X$ must also be zero at that point (denoted as the point of
infinite flatness). By considering that $X$ is an analytic
function, we conclude that infinite flatness may only occur at
$t\to \infty$. Hence if the transition from quintessence to
phantom phase is allowed we must have
\begin{eqnarray}\label{28}
&&\Omega_d^{1\over 2}\cos y\leq {c\over3}(-\lambda_d+\lambda_m-1+\Omega_k), \nonumber\\
&&0<\Omega_d^{1\over 2}\cos y<c.
\end{eqnarray}
Note that the validity of the above inequalities necessitates:
$\lambda_d-\lambda_m+1<\Omega_k$. In the flat case (\ref{28})
becomes
\begin{eqnarray}\label{41}
&&\Omega_d^{1\over 2}\leq {c\over3}(\lambda_m-\lambda_d-1), \nonumber\\
&&0<\Omega_d^{1\over 2}<c,
\end{eqnarray}
hence in this situation, $\lambda_d-\lambda_m+1$ may be only
negative.

In terms of $\Omega_d$, (\ref{28}) may be written as
\begin{equation}\label{29}
\Omega_d<c^2\left(\Omega_k+Min. \{
1,\left({\lambda_m-\lambda_d-1+\Omega_k\over3}\right)^2\}\right),
\end{equation}
which imposes an upper bound on $\Omega_d$ at transition time. At
$w=-1$, we also have
\begin{equation}\label{30}
\Omega_d^{3\over 2}\cos y+{c\over
2}(\lambda_d-\lambda_m+1)\Omega_d+{c\over
2}(\lambda_m-3)(1+\Omega_k)=0.
\end{equation}
In order that transition occurs it is necessary that (\ref{30})
has at least one real root. By considering (\ref{29}) and
(\ref{30}), we arrive at
\begin{equation}\label{31}
{3-\lambda_m\over \Omega_d}<{1\over
1+\Omega_k}\left((\lambda_d-\lambda_m+1)+2Min.\{1,{\lambda_m-\lambda_d-1+\Omega_k\over
3}\}\right).
\end{equation}
This inequality can be written as
\begin{equation}\label{32}
(1+r)(3-\lambda_m)<\gamma,
\end{equation}
or
\begin{eqnarray}\label{33}
&&r<{\gamma\over (3-\lambda_m)}-1,\,\,\,\, if \lambda_m<3\nonumber \\
&&r>{\gamma\over (3-\lambda_m)}-1,\,\,\,\, if \lambda_m>3,
\end{eqnarray}
where we have defined:
$\gamma=(\lambda_d-\lambda_m+1)+2Min.\{1,{\lambda_m-\lambda_d-1+\Omega_k\over
3}\}$. For example if the parameters of the interaction (i.e.
$\lambda_m$ and $\lambda_d$) satisfy: $\lambda_m>3$,
$0<\lambda_m-\lambda_d-1+\Omega_{k0}<3$ and
$(3r_0+2)\lambda_m+\lambda_d-9r_0+2\Omega_{k0}>8$, where $r_0$ and
$\Omega_{k0}$ are the values of $r$ and $\Omega_k$ at transition
time, then (\ref{33}) is satisfied. As another example in a model
characterized by $\lambda_m>3$ ,
$3<\lambda_m-\lambda_d+\Omega_{k0}-1$, and
$(\lambda_m-3)r_0+\lambda_d>0$, the required condition (\ref{33}),
for crossing the $w=-1$ line, is satisfied. E.g. for a closed
universe with $\{\Omega_{k0}=0.02, r_0={3\over 7}\}$, all models
whose interaction parameters satisfy: $\{\lambda_m>3,
7\lambda_d+3\lambda_m>9, \lambda_m-\lambda_d>3.98\}$, fulfill the
condition (\ref{33}).

Note that for $\lambda_m<3$, (\ref{33}) implies
$\gamma>3-\lambda_m$. For negative $\gamma$'s, the second
inequality in (\ref{33}) may be utilized to alleviate the
coincidence problem. Indeed it may pose a positive lower bound on
$r$, in transition epoch.

\section{Conclusion}
In the present paper we have studied the holographic dark energy
model in a closed FRW universe. We have considered an interaction
between (dark) matter and dark energy (see (\ref{7})). By
considering the thermodynamics second law, corresponding to the
entropy assigned to the horizon of the universe (see (\ref{12})),
some relations for relative densities of dark energy ($\Omega_d$),
and dark matter ($\Omega_m$), and geometric parameter ($\Omega_k$)
have been obtained. We have found that in super-accelerated
universe (phantom phase), $r={\Omega_m\over \Omega_d}$, is an
increasing function (see (\ref{21})),  but for accelerated
universe (quintessence phase), depending on the interaction
involved in the theory, $r$ may be a decreasing or an increasing
function of comoving time (see (\ref{9})). We have also shown that
$\Omega_k\over \Omega_d$ is decreasing, provided that the
thermodynamics second law is satisfied (see (\ref{22})). Using the
expression obtained for the equation of state parameter in
(\ref{20}), we have obtained some necessary conditions required
for transition from quintessence to phantom phase (see
(\ref{28})).  These conditions pose some bounds on the dark energy
density at the transition time which can alleviate the coincidence
problem (see (\ref{33})). Note that these bounds depend on the
geometric parameter, $\Omega_k$, as well as on the interaction
parameters.


\begin{thebibliography}{99}
\bibitem{acc}S. Perlmutter et al., Nature (London) 391, 51 (1998);
A. G. Riess et al. (Supernova Search Team Collaboration), Astron.
J. 116, 1009 (1998); S. Perlmutter et al. (Supernova Cosmology
Project Collaboration), Astrophys. J. 517, 565 (1999).
\bibitem{vac}S. Weinberg, Rev. Mod. Phys. 61, 1 (1989); I. Zlatev,
L.-M. Wang, and P. J. Steinhardt, Phys. Rev. Lett. 82, 896 (1999);
V. Sahni, and A. A. Starobinsky, Int. J. Mod. Phys. D 9, 373
(2000); S. M. Carroll, Living Rev. Relativity 4, 1 (2001); T.
Padmanabhan, Phys. Rep. 380, 235 (2003).

\bibitem{cross}V. Sahni, and Y.
Shtanov, J. Cosmol. Astropart. Phys. 11 (2003) 014; U. Alam, V.
Sahni, T. D. Saini, and A. A. Starobinsky, Mon. Not. R. Astron.
Soc. 354, 275 (2004);  V. K. Onemli, and R. P. Woodard, Phys. Rev.
D 70, 107301 (2004); B. Feng, X. L. Wang, and X. M. Zhang, Phys.
Lett. B 607, 35 (2005); D. Huterer, and A. Cooray, Phys. Rev. D
71, 023506 (2005); S. Nesserisa, and L. Perivolaropoulos, Phys.
Rev. D 72, 123519 (2005); S. Nojiri, and S. D. Odintsov, Phys.
Rev. D 72, 023003 (2005); M. Li, B. Feng, and X. Zhang, J. Cosmol.
Astropart. Phys. 12 (2005) 002;  X. Zhang, and F. Wu, Phys. Rev. D
72, 043524 (2005); Z. Guo, Y. Piao, X. Zhang, and Y. Zhang, Phys.
Rev. D 74, 127304 (2006); B. Guberina, R. Horvat, and H. Nikolic,
Phys. Lett. B 636, 80 (2006); H. Mohseni Sadjadi, and M.
Alimohammadi, Phys. Rev. D 74, 043506 (2006); Z. Chang, F. Wu, and
X. Zhang, Phys. Lett. B 633, 14 (2006); G. Cognola, E. Elizalde,
Sh. Nojiri, S. D. Odintsov, and S. Zerbini,
arXiv:hep-th/0611198v3;  M. Alimohammadi, and H. Mohseni Sadjadi,
Phys. Lett. B 648, 113 (2007); M. Bouhmadi-Lopez, and R. Lazkoz,
arXiv:0706.3896v1 [astro-ph]; A. K. Sanyal, arXiv:0710.3486v2
[astro-ph].
\bibitem{inter}L. Amendola, Phys. Rev. D 62, 043511 (2000); W.
Zimdahl, D. Pavon, and L. P. Chimento, Phys. Lett. B 521, 133
(2001); G. Farrar, and P. J. E. Peebles, Astrophys. J. 604, 1
(2004); D. F. Mota, and C. van de Bruck, Astron. Astrophys. 421,
71 (2004); R-G. Cai, and A. Wang, J. Cosmol. Astropart. Phys. 03
(2005) 002; M. Manera, and D. F. Mota, Mon. Not. Roy. Astron. Soc.
371, 1373 (2006); B. Hu, and Y. Ling, Phys. Rev. D 73, 123510
(2006); H. Mohseni Sadjadi, and M. Alimohammadi, Phys. Rev. D 74,
103007 (2006); H. Li, Z. Guo, and Y. Zhang, Int. J. Mod. Phys. D
15, 869 (2006);  E. J. Copeland, M. Sami, and S. Tsujikawa, Int.
J. Mod. Phys. D 15, 1753 (2006); M. Alimohammadi, and H. Mohseni
Sadjadi, Phys. Rev. D 73, 083527 (2006); J. D. Barrow, and T.
Clifton, Phys. Rev. D 73, 103520 (2006); T. Clifton, and J. D.
Barrow , Phys. Rev. D 75 , 043515 (2007); Q. Wu, Y. Gong, A. Wang,
and J. S. Alcaniz, arXiv:0705.1006v3 [astro-ph]; Hao Wei, and S.
N. Zhang; arXiv:0803.3292v3 [astro-ph]; J. He, and B. Wang,
arXiv:0801.4233v2 [astro-ph]; J. Valiviita, E. Majerotto, and R.
Maartens, arXiv:0804.0232v2 [astro-ph]; C. G. Boehmer, G.
Caldera-Cabral, R. Lazkoz, and R. Maartens, arXiv:0801.1565v2
[gr-qc]; K. H. Kim, H. W. Lee, and Y. S. Myung, Phys. Lett. B 632,
605 (2006); M. Quartin, M. O. Calvao, S. E. Joras, R. R. R. Reis,
and I. Waga, arXiv:0802.0546v2 [astro-ph]; N. Pinto-Neto, and B.
M. O. Fraga, arXiv:0711.3602v1 [gr-qc]; C. A. Egan, and C. H.
Lineweaver, arXiv:0712.3099v1 [astro-ph]; K. Karwan, J. Cosmol.
Astropart. Phys. 0805 (2008) 011.
\bibitem{Li1}M. Li, Phys. Lett. B 603, 1 (2004);
Q. Huang, and M. Li, J. Cosmol. Astropart. Phys. 0408 (2004) 013.
\bibitem{holo}S. D. H. Hsu,  Phys. Lett. B 594, 13 (2004); H.
Horvat, Phys. Rev. D 70, 087301 (2004); Q. Huang, and Y. Gong, J.
Cosmol. Astropart. Phys. 0408 (2004) 006; M. Ito, Europhys. Lett.
71, 712 (2005); Y. Gong, B. Wang, and Y. Zhang, Phys. Rev. D 72,
043510 (2005); E. Elizalde, S. Nojiri,  S. D. Odintsov, and P.
Wang,  Phys. Rev. D 71, 103504 (2005); S. Nojiri, and S. D.
Odintsov, Gen. Rel. Grav. 38, 1285 (2006); J. P. Beltran Almeida,
J. G. Pereira, Phys. Lett. B 636, 75 (2006); B. Guberina, R.
Horvat, and H. Nikolic, J. Cosmol. Astropart. Phys. 0701  (2007)
012; Z. Guo, N. Ohta, and S. Tsujikawa, Phys. Rev. D 76, 023508
(2007); X. Zhang, Phys. Rev. D 74 , 103505 (2006);  H. Wei, and S.
N. Zhang, Phys. Rev. D 76, 06003 (2007) 063003; W. Zhao, Phys.
Lett. B 655, 97 (2007); E. N. Saridakis, arXiv:0712.2672v2
[astro-ph]; C. Gao, X. Chen, and Y. Shen, arXiv:0712.1394v3
[astro-ph]; C. Feng, arXiv:0806.0673v1 [hep-th]; J. Lee, H. Kim,
and J. Lee, Phys. Lett. B 661, 67 (2008); Y. S. Myung,
arXiv:0706.3757v2 [gr-qc]; L. Xu, and J. Lu, arXiv:0804.2925v1
[astro-ph]; J. Zhang, X. Zhang, and H. Liu, Phys. Lett. B 659, 26
(2008); S. Lepe, F. Pena, and J. Saavedra, arXiv:0806.0981v1
[gr-qc]; M. Li, C. Lin, and Y. Wang, J. Cosmol. Astropart. Phys.
05 (2008) 023; A. J. M. Medved, arXiv:0802.1753v2 [hep-th]; Y. S.
Myung, and M-G. Seo, arXiv:0803.2913v1 [gr-qc]; H. Kim, J-W. Lee,
and J. Lee, arXiv:0804.2579v2 [gr-qc].
\bibitem{coh}A. Cohen, D. Kaplan and A. Nelson, Phys. Rev. Lett. 82, 4971 (1999).
\bibitem{inf}B. Chen, M. Li, and Y. Wang, Nucl. Phys. B  774, 256 (2007).
\bibitem{abd}B. Wang, Y. Gong, and E. Abdalla, Phys. Lett. B 624, 141
(2005); B. Wang, C. Lin, and E. Abdalla, Phys. Lett. B 637, 357
(2006); B. Wang, J. Zang, C. Lin, E. Abdalla, and S. Micheletti,
arXiv:astro-ph/0607126v3.
\bibitem{me}H. Mohseni Sadjadi, and M. Honardoost, Phys. Lett. B 647, 231 (2007); H.
Mohseni Sadjadi, J. Cosmol. Astropart. Phys. 02 (2007) 026.
\end{thebibliography}
\end{document}